\documentclass[preprint]{aastex}

\usepackage{natbib}
\usepackage{amsmath} \usepackage{amssymb}
\usepackage[colorlinks=true,urlcolor=black,linkcolor=black,citecolor=black]{hyperref}

\shorttitle{From Nonparametric Power Spectra to Inference About Cosmological Parameters}
\shortauthors{Aghamousa, Arjunwadkar, and Souradeep}

 \usepackage{xcolor}
 \newcounter{attnctr} 

 \newcommand{\absval}[1]{\left\lvert #1 \right\rvert}
 
 \newcommand{\sci}[2]{#1 \times 10^{#2}}

\begin{document}

\title{From Nonparametric Power Spectra to Inference About Cosmological Parameters: A Random Walk in the Cosmological Parameter Space}

\author{Amir Aghamousa}
\affil{Centre for Modeling and Simulation, University of Pune, Pune 411 007 India}
\affil{Department of Physics, University of Pune, Pune 411 007 India}
\email{amir@cms.unipune.ac.in}

\author{Mihir Arjunwadkar}
\affil{National Centre for Radio Astrophysics, University of Pune Campus, Pune 411 007 India}
\affil{Centre for Modeling and Simulation, University of Pune, Pune 411 007 India,}
\email{mihir@ncra.tifr.res.in}

\and

\author{Tarun Souradeep}
\affil{Inter-University Centre for Astronomy and Astrophysics, University of Pune Campus, Pune 411 007 India}
\email{tarun@iucaa.ernet.in}

\begin{abstract}
 What do the data, as distinguished from cosmological models, tell us about cosmological parameters that determine the model of the universe?
 In this paper, we address this question in the context of the \textsc{wmap} angular power spectra for the cosmic microwave background (\textsc{cmb}) radiation.
 Nonparametric methods are ideally suited for this purpose because they are model-independent by construction and, therefore, allow inferences that are as data-driven as possible.
 Our analysis is based on a nonparametric fit to the \textsc{wmap} 7-year power spectrum data, with uncertainties characterized in the form of a high-dimensional confidence set centered at this fit.
 For the purpose of making inferences about cosmological parameters, we have devised a sampling method to explore the projection of this confidence set into the space of seven cosmological parameters, namely, $\omega_b, \omega_c, \Omega_\Lambda, H_0, \tau, n_s$, and $A_s$, for a non-flat universe, treating $\Omega_k$ as a derived parameter.
 Our sampling method is justified by its computational simplicity, and validated by the fact that degeneracies and correlations in this cosmological parameter space that are known to be associated with \textsc{cmb} data are correctly reproduced in our results.
 Our results show that
 cosmological parameters are not as tightly constrained by these data alone.
 However, incorporating additional information in the analysis (e.g., constraining the values of $H_0$ or $\Omega_k$) leads to tighter confidence intervals on parameters that are consistent with results based on mainstream parametric methods.
\end{abstract}


\section{Introduction}
\label{introduction}

The angular power spectrum of temperature fluctuations in the cosmic microwave background (\textsc{cmb}) radiation that permeates the universe is one of the most extensively studied data sets in cosmology.
This radiation contains wealth of information about the early universe and its subsequent evolution and,
specifically, about fundamental parameters that govern the universe.
For example, it is well-known that the peaks and dips in the \textsc{cmb} angular power spectrum are directly related to cosmological parameters \citep{HFZ+2001}.
A large number of missions over the past two decades to observe and measure the \textsc{cmb} with ever-increasing angular resolution, precision, and sophistication have led to an accumulation of extensive data for the \textsc{cmb}.
The \textsc{wmap} mission, e.g., has so far released four data realizations representing cumulative observations at the end of 1, 3, 5, and 7 years of operation.
Typical \textsc{cmb} data analysis pipelines \citep{TG2007} begin by constructing sky maps from time-ordered observations
and applying corrections for contamination by the foreground.
\textsc{cmb} data analysis typically ends with a regression step that estimates the angular power spectrum
together with inferences about the spectrum and the cosmological parameters under consideration.

The essential difficulty in inferring cosmological parameters from an estimated \textsc{cmb} angular power spectrum is that the inferential entities,
such as likelihood functions, posterior distributions, or the nonparametric confidence set to be described below,
belong to one space (the {power spectrum space, or PS-space}), whereas the inferred entities (i.e., cosmological parameters)
belong to a different space (the {cosmological parameter space, or CP-space}),
and the mapping available is a one-way mapping that maps parameters to spectrum.
That is, given an appropriate set of cosmological parameter values, it is generally possible to compute or approximate the
corresponding \textsc{cmb} angular power spectrum through the solution of a Boltzmann equation \citep{Seljak:1996is,PY1970,BE1984,BE1987,dodelson2003modern}.
This approach, with variations and embellishments geared for accuracy and computational efficiency,
is at the heart of ubiquitous computational tools such as
\textsc{cmbfast} \citep{Seljak:1996is}, \textsc{camb} \citep{Lewis:1999bs}, etc.
In other words, it is possible to computationally map a set of cosmological parameters onto a \textsc{cmb} power spectrum.
However, the reverse of this mapping that would map a power spectrum back to a set of cosmological parameters is not computable directly, let alone in an efficient manner.
Hence, this inverse mapping problem needs to be formulated as a computational problem involving search or sampling.
Mainstream methods based on maximum likelihood or maximum posterior approaches typically resort to sampling via Markov chain Monte Carlo methods (available in the form of tools such as \textsc{cosmomc} \citep{Lewis:2002ah}) to circumvent the inverse mapping problem.

In our previous work \citep{AAS2012}, we estimated the \textsc{cmb} angular power spectrum from the four \textsc{wmap} data realizations using a nonparametric function estimation methodology \citep{Beran2000,GMN+2004}.
This methodology does not impose any specific form or model for the power spectrum,
and determines the fit by optimizing a measure of smoothness that depends only on characteristics of the data. 
This ensures that the fit and the subsequent analysis is approximately model-independent for large data size and, therefore, is as data-driven as it could possibly be.
Further, this methodology quantifies the uncertainties in the fit in the form of a high-dimensional ellipsoidal confidence set that is centered at this fit and captures the true but unknown power spectrum with a pre-specified probability.
This confidence set is the prime inferential object of this methodology.

By construction, this nonparametric confidence set (a) is centered at the nonparametric fit,
and (b) ensures that the true but unknown power spectrum is contained within itself with a probability in excess of a pre-specified threshold (called the \emph{confidence level}) of $(1-\alpha)$.
Further, inferences about any number of features of the estimated power spectrum are simultaneously valid at the same level of confidence.
At a given level of confidence, the boundary of this confidence set partitions the PS-space into two parts:
All spectra inside this confidence set are considered equally likely candidates for the unknown truth,
whereas all spectra outside are rejected as candidates for the unknown truth, with $\alpha$ as the probability of an incorrect rejection.

Mainstream methods used for estimating the \textsc{cmb} power spectrum as well as for making inferences about cosmological parameters (see, e.g., \citet{LDH+2011}) are model-based, by and large Bayesian in their outlook, and do not share these unique inferential features (see \citet{GMN+2004} for a thorough discussion).
On general grounds, one could argue \citep{AAS2012} that such powerful nonparametric methods could be used to validate inferences made using parametric, model-based methods which involve stronger assumptions about the data or the true but unknown power spectrum.

\begin{figure}
 \centering
 \includegraphics[width=0.9\textwidth]{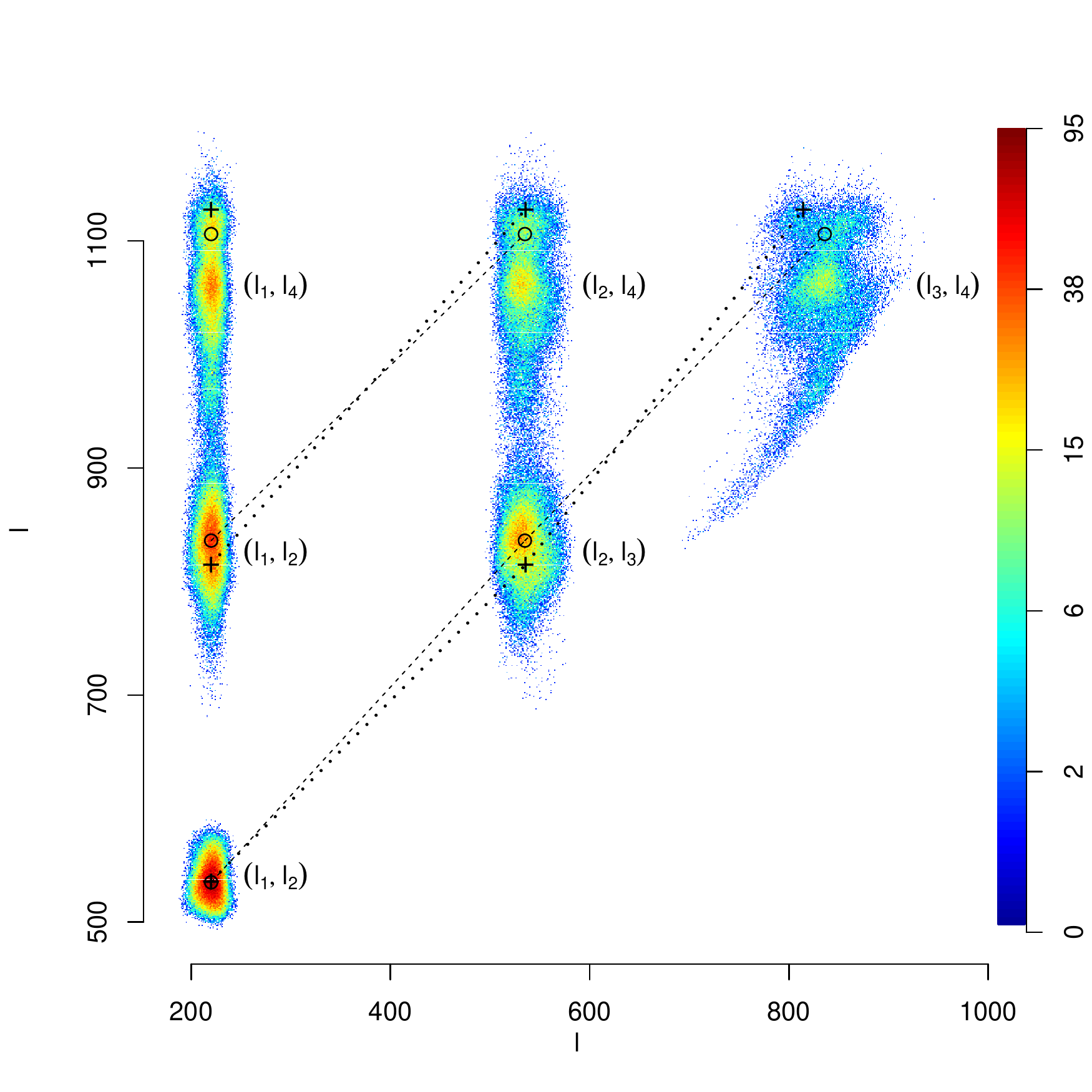}
 \caption{\label{fig:harmonicity} Nonparametric indication of the harmonicity of peaks in the CMB angular power spectrum as inferred from the \textsc{wmap} 7-year data. This is a 2-dimensional color-coded histogram for the $(l_i,l_j)$ scatter, where $l_i$ is the position of the $i$th peak, for a sample of 68000 4-peaked spectra from the 95\% $(2\sigma)$ confidence set. Color indicates log(count) over an $l$-grid with unit spacing in both directions. We interpret the arrangement of the six peaks in this histogram over a regular rectangular grid as nonparametric evidence for the basic physics of harmonicity of acoustic oscillations of the baryon-photon fluid that gave rise to CMB anisotropies. Also indicated are the peak-peak locations for the $\Lambda$CDM parametric fit \citep{LDH+2011} (crosses), and those for our nonparametric fit (circles), connected by dotted and dashed lines respectively, as a guide to the eye.}
\end{figure}

Our previous work \citep{AAS2012} focused on estimating the \textsc{cmb} power spectrum from \textsc{wmap} data realizations, and on making inferences about features of the true but unknown angular power spectrum.
For example, our results already indicated the basic physics of harmonic features in the CMB power spectrum in a model-independent manner; we illustrate these harmonic features in Fig.\ \ref{fig:harmonicity} for the \textsc{wmap} 7-year data.
In this paper, we extend this work to making inferences about cosmological parameters using the nonparametric confidence set for the \textsc{wmap} 7-year data.
The cosmological parameters we consider here,
following \citet{LDH+2011},
are $(\omega_b, \omega_c, \Omega_\Lambda, H_0, \tau, n_s, A_s )$,
together with $\Omega_k$ as a derived parameter, for a non-flat universe without a massive neutrino component (i.e., $\Omega_\nu = 0$).
Our objective is to see how data-driven, nonparametric uncertainties on cosmological parameters compare with, and perhaps validate (or otherwise), mainstream parametric results.
For this purpose, we devise a Markov chain Monte Carlo method that samples the CP-space in such a way that the resulting density over the PS-space is,
at least in principle, uniform within the cosmologically meaningful region of the confidence set around our nonparametric fit.
A somewhat complex method \citep{BSM+2007} for mapping the confidence set boundary in the PS-space into the CP-space does exist.
Recently, this method has also been adopted \citep{DCS2012} for mapping contours of a parametric likelihood function into the CP-space.
Our method, on the other hand, is appealing for its conceptual and computational simplicity,
is by and large at least as computationally efficient as this method for \textsc{cmb} data sets such as the \textsc{wmap} 7-year data
and, in some cases, appears to outperform this method for the \textsc{wmap} data sets.

We apply our method to finding confidence intervals for the cosmological parameters mentioned above.
We obtain numerical glimpses of their pairwise joint distributions which show the well-known degeneracies and correlations in
the CP-space resulting from the fact that \textsc{cmb} data alone does not constrain all cosmological parameters sufficiently precisely.
This result also serves as a validation of our method.
The nonparametric, model-independent, data-driven uncertainties on cosmological parameters thus obtained
turn out to be much larger than the parametric confidence intervals reported by mainstream methods (see, e.g., \citet{LDH+2011}).
This suggests that parametric confidence intervals should be interpreted with adequate caution.
However, supplementing our nonparametric analysis with additional/prior information about cosmological parameters produces much tighter confidence intervals.
Our results thus validate mainstream parametric results with additional information.

In what follows, Sec.\ \ref{method} presents our method for sampling the projection of a nonparametric confidence set in the CP-space.
Results are presented in Sec.\ \ref{results}.
We conclude this paper with a discussion in Sec.\ \ref{discussion}.

\section{Methodology}
\label{method}

\subsection{An Overview of the Nonparametric Confidence Set}
\label{method:npcs}

We begin with a brief overview of the nonparametric regression methodology and the confidence set construct.
This overview is based on \citep{AAS2012};
similar overviews have also appeared in \citep{BSM+2007,GMN+2004}.
Additional details can be found in \citep{Beran2001,Beran2000}.

The \textsc{cmb} angular power spectrum data are assumed to be of the form
\begin{equation}
 Y_l = C_l + \epsilon_l,
 \label{eq:splusn}
\end{equation}
consisting of $N$ data points observed over multipole range $l_{min} \le l \le l_{max}$,
where $C_l$ is the value of the true but unknown power spectrum $C$ at $l$, which is to be estimated from data.
The noise variables $\epsilon_l$ are assumed to have a mean-0 normal distribution with a known covariance matrix $\Sigma$.
In the description of the regression method and the confidence set below, we refer to $C$ and $l$ as $f$ and $x$ respectively.

The nonparametric regression method we use is based on expanding the unknown regression function $f$,
assumed to be square-integrable, in a complete orthonormal basis $\{ \phi_j(x), j = 0, 1, \ldots \}$ as
$f(x) = \sum_{j=0}^\infty \beta_j \phi_j(x).$
The estimator $\widehat{f}$ of $f$ is taken to be of the form
$\widehat{f}(x) = \sum_{j=0}^{N-1} \widehat{\beta}_j \phi_j(x),$
where
$\widehat{\beta}_j = \lambda_j N^{-1} \sum_{i=0}^{N-1} Y_i \phi_j(x_i)$.
The \emph{shrinkage parameters}
$1 \ge \lambda_0 \ge \lambda_1 \ge \ldots \ge \lambda_{N-1} \ge 0$
are obtained by minimizing an inverse-noise-weighted risk function that attempts to balance the bias of $\widehat{f}$ with its variance to achieve optimal smoothness in the fit.

A $(1 - \alpha)$ confidence set around the fit $\widehat{f}$ is defined as follows.
A $(1 - \alpha)$ confidence set centered at the estimated coefficient vector $\widehat{\beta}$ is given by
\begin{equation}
 \label{eq:csetD}
 \mathcal{D}_{N,\alpha} = \left\{ \beta: (\beta-\widehat{\beta})^T W (\beta-\widehat{\beta}) \le {r}^2_\alpha \right\},
\end{equation}
where the quantity ${r}_\alpha$, called the \emph{confidence radius}, is a monotonically increasing function of $(1 - \alpha)$,
and the matrix $W$ is related to the inverse noise weighting; see \citep{AAS2012} for details.
The corresponding confidence set on the true regression function $f$ is given by
\begin{equation}
 \label{eq:csetB}
 \mathcal{B}_{N,\alpha} = \left\{ f(x) = \sum_{j=0}^{N-1} \beta_j \phi_j(x) : \beta \in \mathcal{D}_{N,\alpha} \right\}.
\end{equation}
Confidence intervals for any functional of the fit $\widehat{f}$ (e.g., cosmological parameters, heights and locations of peaks and dips, etc.)
are obtained by recording the extreme values of the functional over this confidence set.
Such frequentist confidence intervals are interpreted (paraphrasing from \citet{BSM+2007}; see also \citet{Wasserman2004}) as follows:
When applied to a series of data sets, a frequentist confidence interval traps, by construction, the true value of a parameter for at least $100 ( 1 - \alpha )\%$ of the data sets.

The method provides the formal guarantees that
(a) all such confidence intervals on any number of functionals are simultaneously valid at the same level of confidence $(1-\alpha)$ and, moreover,
(b) this confidence set contains the true but unknown function $f$ with probability $\ge (1 - \alpha)$.
Therefore, the probability of an incorrect rejection (i.e., rejection of an outside point as a candidate for the true $f$ even if it happens to be the true $f$ itself) is $\alpha$.
Further, any prior information about the true spectrum or its features, or cosmological parameters, can in principle be incorporated in this approach by considering an appropriate subset of the full confidence set.

\subsection{Mapping the Confidence Set into the CP-space via Sampling}
\label{method:mcmc}

We interpret the latter guarantee (b) (Sec.\ \ref{method:npcs})
as saying that the probability of finding the true but unknown function $f$ is distributed uniformly inside the confidence set.
The sampling method we develop here is therefore designed to sample the inside of the confidence set
in the PS-space with a uniform density.
We use \textsc{camb} \citep{Lewis:1999bs} to compute the lensed power spectrum given a set of cosmological parameters.
We also make two additional parametric assumptions; namely,
the power-law form of the primordial power spectrum,
and standard values
(obtained from the \textsc{wmap} 7-year data page {\scriptsize \url{http://lambda.gsfc.nasa.gov/product/map/dr4/params/lcdm_sz_lens_wmap7.cfm}})
for other parameters that determine the power spectrum.
Practically, the use of \textsc{camb} limits our sampling method to regions of the confidence set that are accessible via \textsc{camb}.
The corresponding density in the CP space is determined by the nature of the unknown inverse mapping from the PS- to the CP-space,
and is expected to be highly non-uniform.
Our method consists of two separate Markov chain Monte Carlo (MCMC) stages.

\paragraph{Stage 1: Locating a CP-space point whose spectrum lies inside the confidence set.}
Starting from an arbitrary or randomly chosen point $p_0$ in a cosmologically plausible area of an $M$-dimensional CP-space,
we first locate a point in the CP-space whose power spectrum lies on or inside the confidence set in the PS-space.
We do this by setting up a Markov chain via the Metropolis algorithm \citep{Wasserman2004}
to sample from a probability density function $g$, called the \emph{guide density}, that is centered
at the center $\widehat{C}$ of the nonparametric confidence set (Eq.\ \ref{eq:csetB}):
\begin{equation}
 g( p ) \propto \exp\left( - { d^2( C(p), \widehat{C} ) \over 2 b^2 }\right),
 \label{eq:gofd}
\end{equation}
where
\begin{equation}
 d(C,C^\prime) = \sqrt{ {1 \over N} \sum_{l=l_{min}}^{l_{max}} \left( C_l - C^\prime_l \over \sigma_l \right)^2  }
 \label{eq:nwd}
\end{equation}
is the noise-variance-weighted distance between the power spectra $C, C^\prime$;
$C(p)$ stands for the power spectrum for a parameter vector $p$;
$\sigma_l$ is the (known) standard deviation of noise in the data (Eq.\ \ref{eq:splusn}) at $l$;
and the scale parameter $b$ is chosen so that the guide density $g$ has a nonzero, numerically finite value at the initial point $p_0$.
At the $k$-th step of the Metropolis algorithm, a move from the current location $p_k$ to $p$
is proposed using the proposal density
\begin{equation}
 q( p | p_k ) \propto \exp\left( -{ 1 \over 2 }( p - p_k )^T A ( p - p_k ) \right),
 \label{eq:proposal}
\end{equation}
and accepted with probability
\begin{equation}
 a( p | p_k ) = \min\left\{ 1, { g( p ) \over g( p_k ) } \right\}.
 \label{eq:acceptance}
\end{equation}
The matrix $A = \mbox{diag}( 1/a_1^2, \ldots, 1/a_M^2 )$ in the proposal density is defined in terms of step sizes $a_1, \ldots, a_M$ along the $M$ directions in the CP-space:
We use the arbitrary but reasonable prescription $a_i = 0.01 \times \mbox{the}$ allowed range for the $i$-th cosmological parameter.
This MCMC scheme thus performs a Gaussian random walk in the CP-space that is directed towards the parameter vector corresponding to the center $\widehat{C}$ of the confidence set.
We terminate this chain as soon as a point on or inside the confidence set is located.

\paragraph{Stage 2: Sampling the confidence set with uniform density.}
Starting from a point in the CP-space that corresponds to a power spectrum on or inside the confidence set,
we set up another Markov chain to sample the confidence set with a uniform density.
For this chain, the guide density $g$ is now replaced by a uniform density over the confidence set in the PS-space.
The proposal density and our prescription for the step size parameters $a_i$ remains the same as that in the previous stage;
however, the step size parameters $a_i$ may also be monitored and adjusted by trial-and-error so as to keep the average acceptance ratio between acceptable limits.
Since the sampling density is now uniform over the confidence set, the acceptance/rejection step of the Metropolis algorithm takes a particularly simple form:
If the proposed parameter vector $p$ corresponds to a power spectrum inside or on the confidence set, accept it; else, reject it.

\paragraph{Ensuring adequate sampling of the confidence set.}
MCMC methods in general, and the Metropolis algorithm in particular, are known to get stuck in local maxima in the probability density function being sampled \citep{BSM+2007}.
To circumvent such problems in the stage 2 above, we run an arbitrary number (typically 100) of chains starting from arbitrary points on or inside the confidence set, and pool them together.
This is justified because we are primarily interested in the boundaries of the confidence set as projected in the CP-space,
and not in the density variations within these boundaries.
We assess convergence heuristically by pausing all chains every 1000 steps or so, pooling them together, and checking if the resulting pairwise scatterplots have changed substantially.
As an additional diagnostic check for adequate sampling, we run stage 1 above from another set of random starting points in the CP-space, and see if they end up in a new region of the confidence set that has not been sampled before.
From preliminary results, we also look for indications (such as sparse disconnected patches) of undersampling of the CP-space.
To improve sampling of such regions, we typically resample these regions by restricting our parameter search ranges to these regions.
In the end, we pool all these chains together.

\section{Results}
\label{results}

The confidence set we use in this work is centered at the monotone nonparametric fit \citep{AAS2012} to the \textsc{wmap} 7-year data.
The dimensionality of the confidence set is 1199, whereas the fit belongs to a subspace with approximately 103 effective degrees of freedom.
The 95\% ($2\sigma$), 68\% ($1\sigma$), 38\% ($0.5\sigma$), and 20\% ($0.25\sigma$) confidence radii are
$0.35509, 0.28696, 0.23154$, and $0.18355$ respectively.
This nonparametric fit \citep{AAS2012} and the parametric fit \citep{LDH+2011} to the \textsc{wmap} 7-year data are quite close (see Fig.\ 4 in \cite{AAS2012}):
Their average relative difference
$$
 {1 \over (l_{max}-l_{min}+1)} \sum_{l=l_{min}}^{l_{max}} \absval{{\widehat{C}_l-\tilde{C}_l \over \tilde{C}_l}} \approx 0.09,
$$
where $\widehat{C}$ and $\tilde{C}$ are the nonparametric and the parametric fits respectively.

\subsection{Cosmological Parameters}

Fig.\ \ref{fig:prior-nill} shows nonparametric pairwise scatterplots
without any additional/prior information used while sampling the confidence set,
for the cosmological parameters $(\omega_b,\omega_c,\Omega_\Lambda,\Omega_k,H_0,\tau,n_s,A_s)$.
These scatterplots represent four levels of confidence;
namely, 95\% ($2\sigma$, black), 68\% ($1\sigma$, gray), 38\% ($0.5\sigma$, blue), 20\% ($0.25\sigma$, red).
$\Omega_k$ is a derived parameter here, obtained as $\Omega_k = 1 - ( \Omega_b + \Omega_c + \Omega_\Lambda )$,
assuming that the massive neutrino component $\Omega_\nu = 0$.
We see that all well-known degeneracies and correlations characteristic of \textsc{cmb} data
(e.g., correlations between the $(\Omega_k,H_0)$ and $(\Omega_k,\Omega_\Lambda)$ pairs)
are correctly reproduced by our method.
Our nonparametric confidence intervals (Table \ref{tab:ci}, column 2), defined by the extreme values for each parameter in this scatter, are much wider than the corresponding parametric ones (\citet{LDH+2011}; see also our Table \ref{tab:pe}).
This point has been discussed further in Sec.\ \ref{discussion}.

\begin{figure}
 \centering
 \begin{minipage}{0.48\textwidth}
 \includegraphics[width=\textwidth]{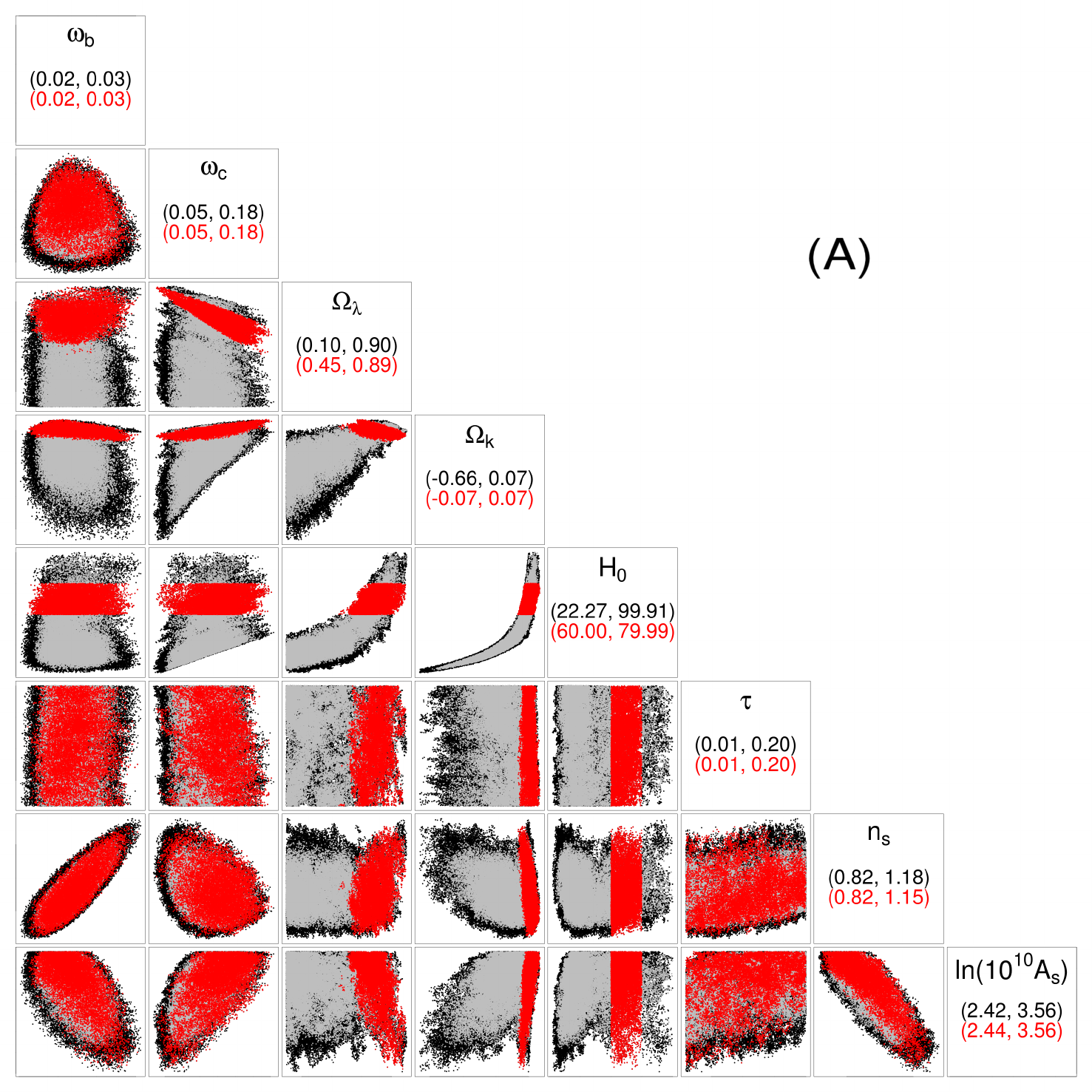}
 \end{minipage}
 \begin{minipage}{0.48\textwidth}
 \includegraphics[width=\textwidth]{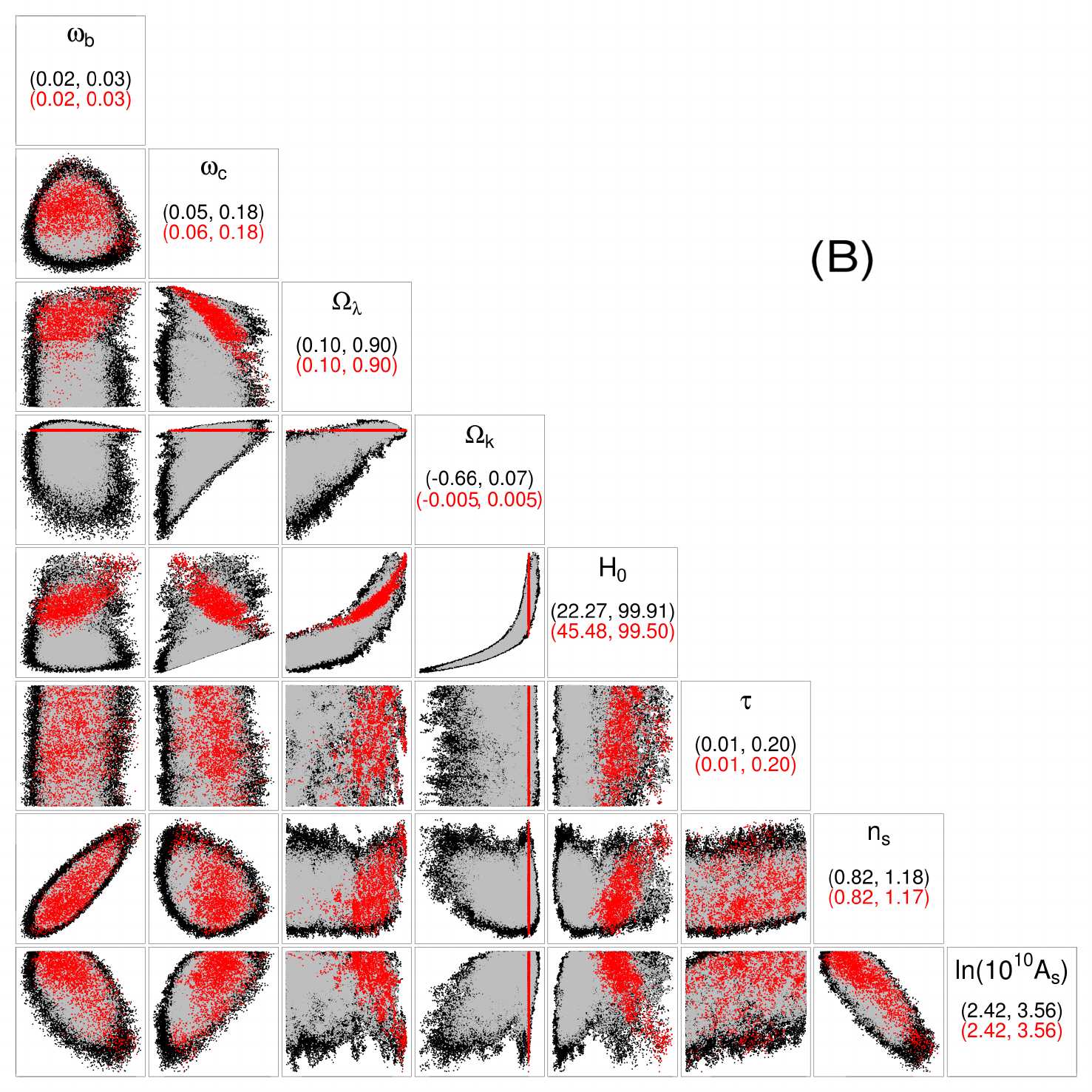}
 \end{minipage}
 \caption{\label{fig:prior-H0-Ok} Nonparametric pairwise scatter (red) of cosmological parameters with additional/prior information, namely, $60 \le H_0 \le 80$ (A) and $-0.005 \le \Omega_k \le +0.005$ (B). Also shown is the full nonparametric scatter at two levels of confidence; namely, 95\% ($2\sigma$, black) and 68\% ($1\sigma$, gray). Notice that the confidence intervals have shrunk considerably after constraining $H_0$ or $\Omega_k$. Panels on the diagonal list the range of values for the corresponding parameter.}
\end{figure}

Inclusion of additional/prior information in our sampling, however, tends to shrink the confidence intervals.
Fig.\ \ref{fig:prior-H0-Ok} shows the nonparametric pairwise scatterplots (red) after constraining, respectively, $H_0$ to the range $60 \le H_0 \le 80$, and $\Omega_k$ to the range $-0.005 \le \Omega_k \le +0.005$.
Also shown is the full nonparametric scatter at two levels of confidence; namely, 95\% ($2\sigma$, black) and 68\% ($1\sigma$, gray).
Notice that the confidence intervals (Table \ref{tab:ci}, columns 3 and 4) have shrunk considerably with this additional information.

\begin{figure}
 \centering
 \includegraphics[width=0.9\textwidth]{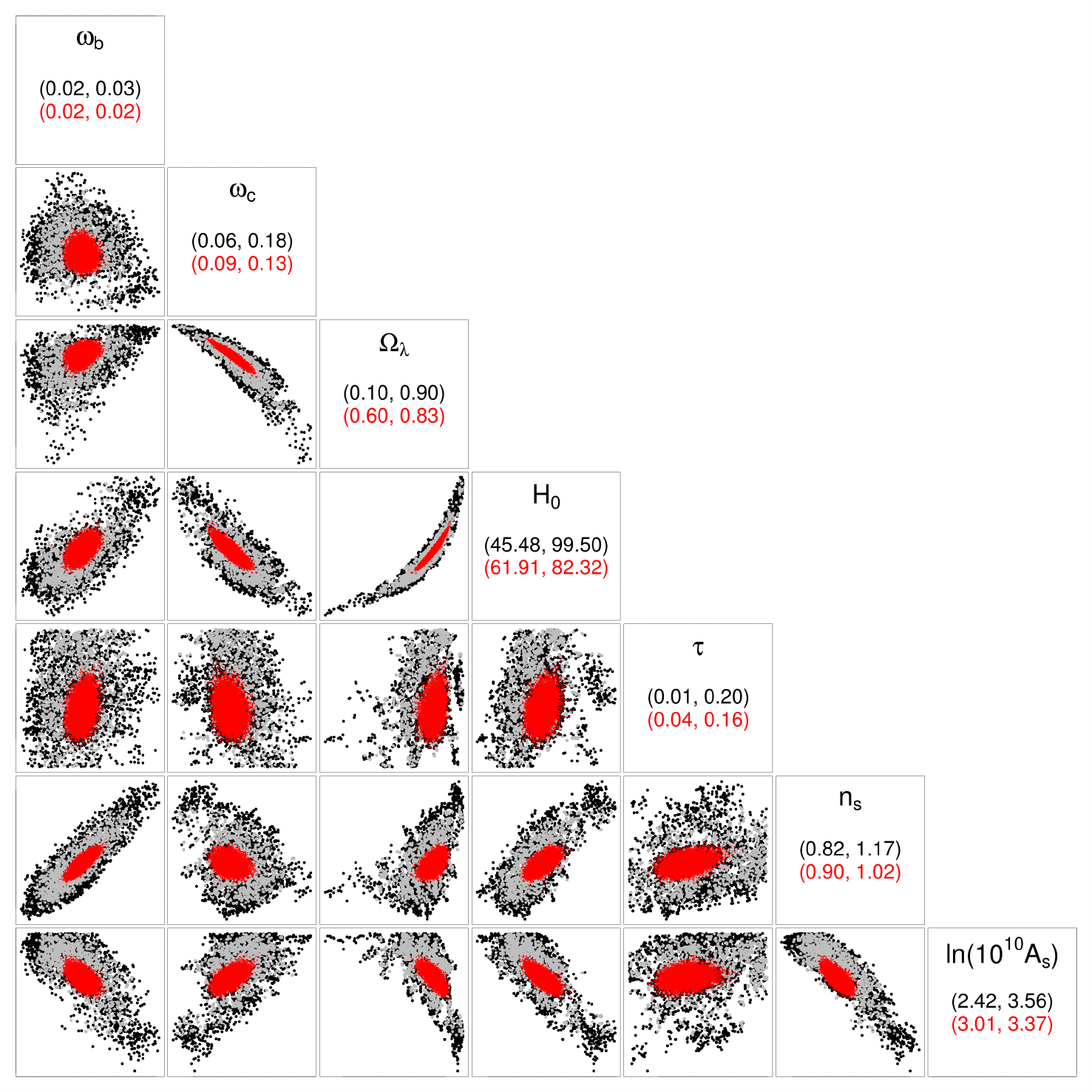}
 \caption{\label{fig:p-vs-np}
 Parametric scatter (red) for a Markov chain Monte Carlo sample from a parametric likelihood function \citep{LDH+2011} for the \textsc{wmap} 7-year data, under the assumption of a flat universe ($\Omega_k = 0$). Also indicated for reference are the nonparametric pairwise scatters of cosmological parameters at the 95\% ($2\sigma$, black) and 68\% ($1\sigma$, gray) confidence levels with the restriction $-0.005 \le \Omega_k \le 0.005$ (Fig.\ \ref{fig:prior-H0-Ok}B). Panels on the diagonal list the range of values for the corresponding parameter. We see that the parametric scatter is more or less contained withing the nonparametric scatter. Extreme parameter values in the parametric scatter (red) are listed in Table \ref{tab:pe}.}
\end{figure}

In Fig.\ \ref{fig:p-vs-np}, we overlay our nonparametric scatter (black) at 95\% ($2\sigma$) confidence level
with a parametric scatter (red) representing a Markov chain Monte Carlo sample from the parametric likelihood function for the \textsc{wmap} 7-year data \citep{LDH+2011}.
For this comparison, we have set $\Omega_k=0$ as this Markov chain assumes a flat universe.
The parametric scatter is more or less centered within the nonparametric scatter, which we interpret as a nonparametric validation of the parametric result.

Fig.\ \ref{fig:cosmic-triangle} is a nonparametric cosmic triangle plot \citep{BOP+1999} at 95\% $(2\sigma)$ confidence level,
with no additional information (black scatter), as well as with the restriction $60 \le H_0 \le 80$ (red scatter).
The right-hand panel in this figure indicates the possible cosmologies in different parts of the cosmic triangle.

\begin{figure}
 \centering
 \includegraphics[width=0.9\textwidth]{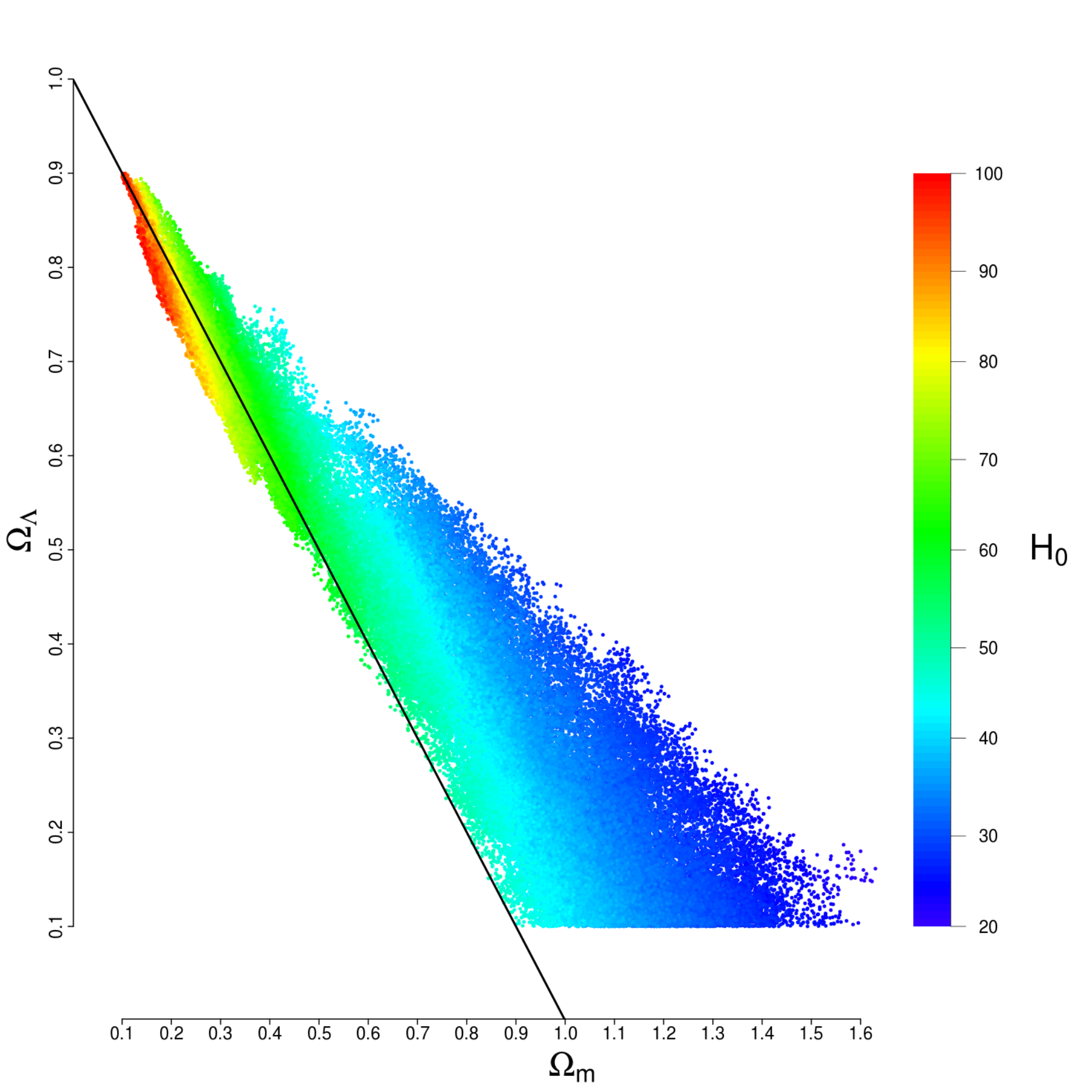}
 \caption{\label{fig:Ov-vs-Om} The nonparametric $\Omega_\Lambda$ vs.\ $\Omega_m$ scatter at the 95\% ($2\sigma$) confidence level, color-coded for the Hubble parameter $H_0$. The black line corresponds to a flat universe (assuming that the massive neutrino component $\Omega_\nu=0$). This figure may be compared with Fig.\ 14 in \citep{LDH+2011}. Our nonparametric results indicate, consistent with \citep{LDH+2011}, that an open universe is more or less ruled out by the \textsc{wmap} 7-year data. However, a closed universe cannot be ruled out by the current \textsc{cmb} data alone, unless additional information (e.g., about $H_0$) is incorporated in the analysis.}
\end{figure}

\begin{figure}
 \centering
 \includegraphics[width=0.9\textwidth]{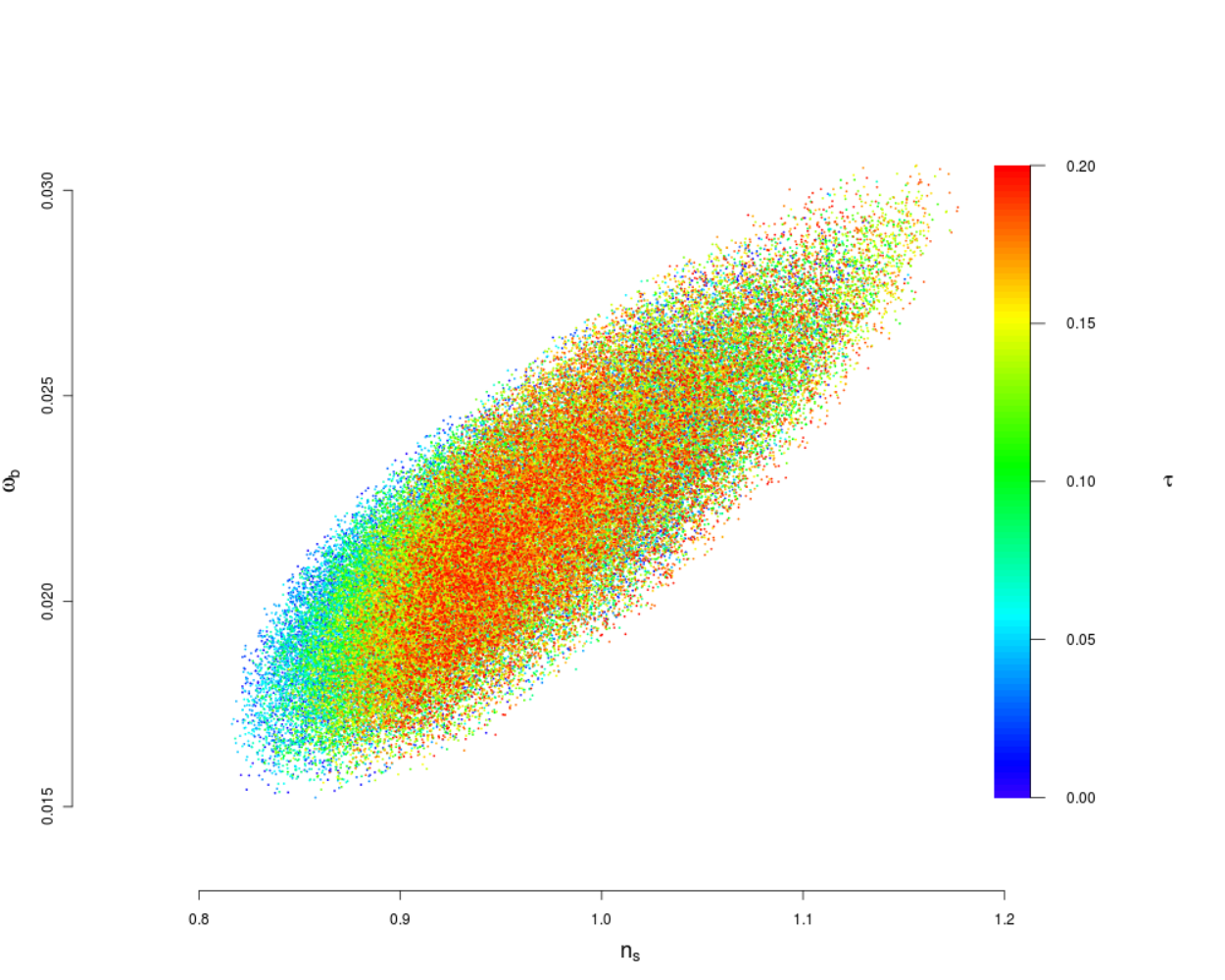}
 \caption{\label{fig:ob-vs-ns} The nonparametric $\omega_b$ vs.\ $n_s$ scatter at the 95\% ($2\sigma$) confidence level, color-coded for the optical depth parameter $\tau$. This figure may be compared with Fig.\ 2.7 in \citep{TPC2006}. This is a nonparametric illustration of the well-known degeneracy between these three parameters. This degeneracy may be broken by supplementing CMB anisotropy data with CMB polarization data.}
\end{figure}

Fig.\ \ref{fig:Ov-vs-Om} and \ref{fig:ob-vs-ns} illustrate two of the well-known geometrical degeneracies
(see, e.g., \citet{TPC2006,SAA+2001}) in the CMB data, but arrived at through our methodology.
Specifically, Fig.\ \ref{fig:Ov-vs-Om} shows the nonparametric scatter in the $\Omega_\Lambda$--$\Omega_m$ plane at 95\% $(2\sigma)$ confidence level,
color-coded for the Hubble parameter $H_0$.
The black line in this figure corresponds to a flat universe (assuming that the massive neutrino component $\Omega_\nu=0$).
This figure may be compared with Fig.\ 14 in \citet{LDH+2011}:
Both these figures indicate, consistent with what is known in the field, that an open universe is more or less ruled out by the \textsc{wmap} 7-year data.
However, a closed universe cannot be ruled out by the current \textsc{cmb} data alone, unless additional information (e.g., about $H_0$) is incorporated in the analysis.

Fig.\ \ref{fig:ob-vs-ns} similarly shows the nonparametric scatter in the $\omega_b$--$n_s$ plane at 95\% $(2\sigma)$ confidence level, color-coded for the optical depth $\tau$.
This figure may be compared with Fig.\ 2.7 in \citep{TPC2006}.
As is well-known, the degeneracy between the spectral index $n_s$ and optical depth $\tau$ is associated with reionization, and cannot be broken with CMB anisotropy data alone \citep{SAA+2001}.
Additional information, e.g., coming from accurate CMB polarization data, or through tighter constraints on other cosmological parameters such as $\omega_b$, will help determine the $n_s$ and $\tau$ parameters with greater accuracy.
Our results lead to the same conclusion from a data-driven, model-independent viewpoint.

\subsection{Performance of the Method}

\begin{figure}
 \centering
 \includegraphics[width=0.9\textwidth]{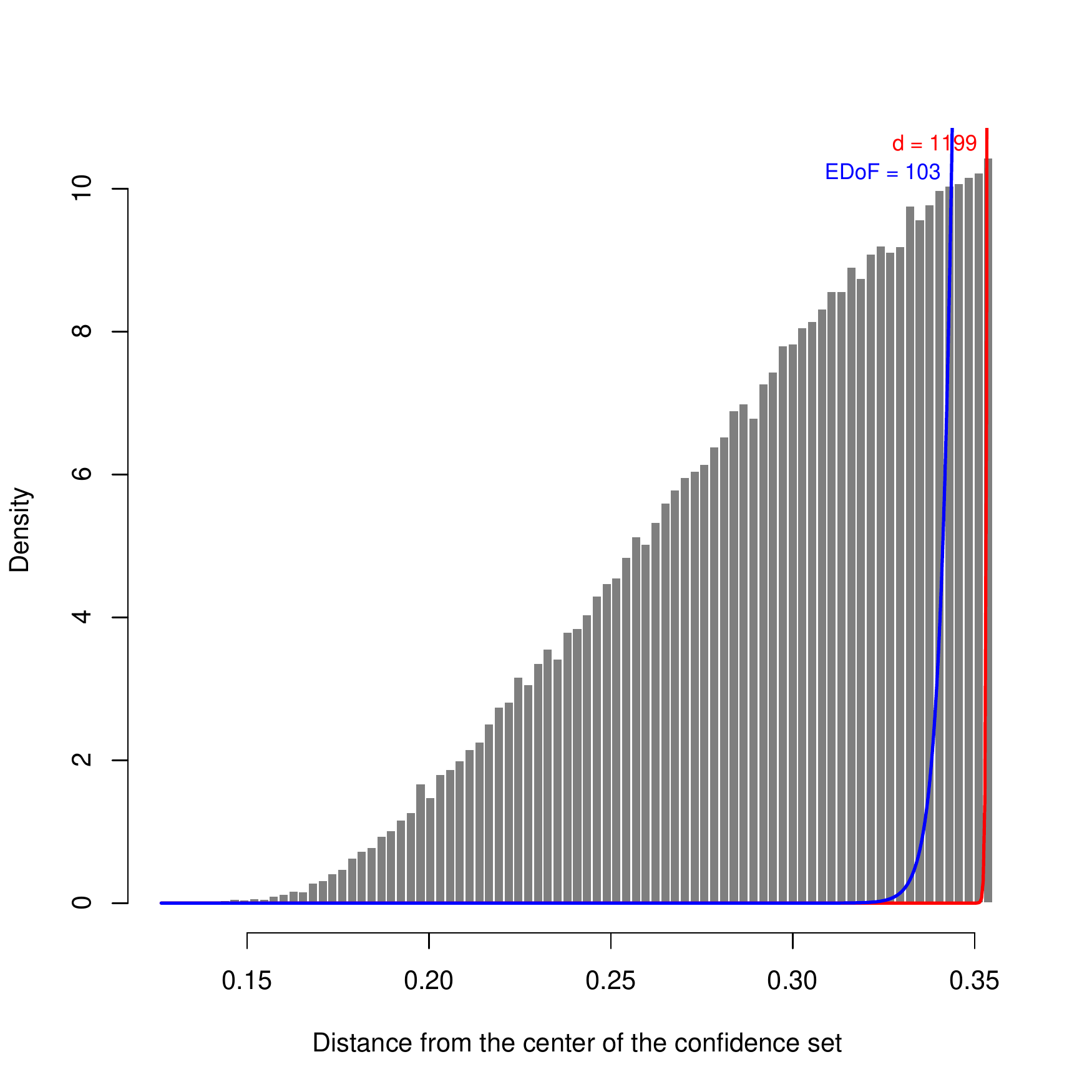}
 \caption{\label{fig:disthist} Distribution of distances from the center of the confidence set for a sample generated using our sampling method. The two curves represent theoretical distance distributions of the form $d r^{d-1} / r_\alpha^d$ corresponding to uniform sampling of $d$-dimensional spheres ($d=103$: blue, $d=1199$: red). Here, 1199 is the full dimension of the nonparametric fit, whereas 103 is the approximate number of its effective degrees of freedom \citep{AAS2012}, and $r_\alpha$ is the radius of the $(1-\alpha)$ confidence set ($\alpha=0.05$). The discrepancy between these curves and the histogram originates in the fact that not all power spectra that belong to the confidence set are cosmologically plausible and computationally reachable via \textsc{camb}.}
\end{figure}

\paragraph{Distance distribution.}
Fig.\ \ref{fig:disthist} shows the distribution of distances from the center of the confidence set of power spectra sampled using our method.
Also plotted are the theoretical distance distributions of the form $d r^{d-1} / r_\alpha^d$ for a $d$-dimensional spherical set sampled with uniform density, where $r_\alpha$ is the radius of the $(1-\alpha)$ confidence set.
The dimensions that are relevant here are $d=1199$ and $d \approx 103$, corresponding to the full dimension of the nonparametric fit and its effective degrees of freedom \citep{AAS2012} respectively.
The reason for the discrepancy between either of the theoretical distributions and the histogram of the sampled distances is that not every power spectrum that belongs to the confidence set is cosmologically meaningful or computationally accessible via \textsc{camb}.

\begin{figure}
 \centering
 \begin{minipage}{0.48\textwidth}
 \includegraphics[width=\textwidth]{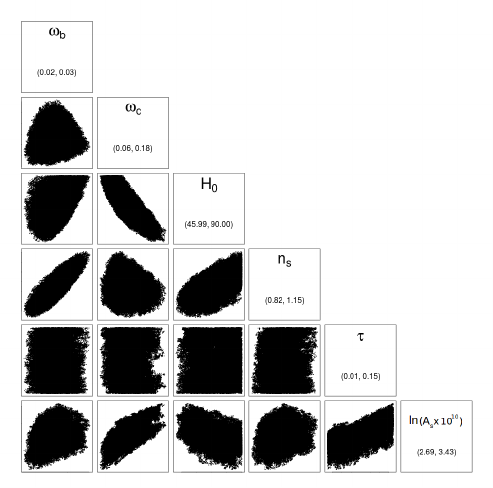}
 \end{minipage}
 \begin{minipage}{0.48\textwidth}
 \includegraphics[width=\textwidth]{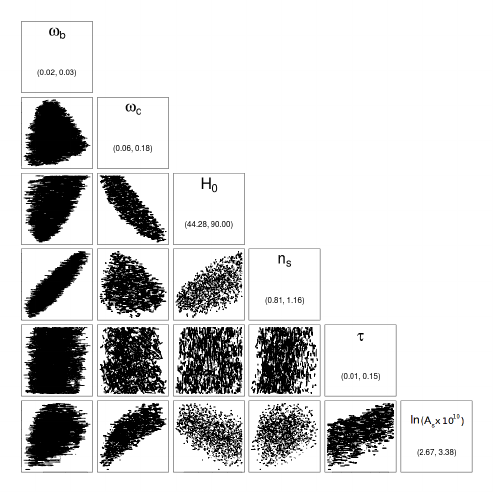}
 \end{minipage}
 \caption{\label{fig:aps-comparison} Comparison of results obtained using our method (left) and the APS \citep{DCS2012} (right) for sampling the wmap 7-year nonparametric confidence set. Number of power spectra sampled from within this confidence set by both methods is approximately 100,000. For generating these 100,000 samples, our method sampled a total of about 142,000 spectra (70\% efficiency), whereas the \textsc{aps} required about 200,000 samples (50\% efficiency). Panels on the diagonal list the range of values for the corresponding parameter. Following \citep{DCS2012}, we set $\Omega_k=0$ for the purpose of this comparison.}
\end{figure}

\paragraph{Comparison with an alternate method.}
A method based on kriging and straddling \citep{BSM+2007} attempts to map boundaries of nonparametric confidence sets in the PS-space back into the CP-space.
This method has recently been adopted, in the form of a tool called \textsc{aps} \citep{DCS2012}, for locating CP-space loci corresponding to PS-space contours of a likelihood function.
Fig.\ \ref{fig:aps-comparison} provides a comparison between \textsc{aps} and our sampling method for the default \textsc{aps} parameter set $(\omega_b,\omega_c,H_0,\tau,n_s,A_s)$, together with $\Omega_k=0$, where both methods sample the same confidence set for \textsc{wmap} 7-year data.
The bounds used by \textsc{aps} for these cosmological parameters are typical and tight.
The number of power spectra sampled by either method from inside the confidence set was $\approx 100,000$.
In Fig.\ \ref{fig:aps-comparison}, we see that the results obtained by the two methods are qualitatively equivalent.
On the other hand, for these bounds, the acceptance ratio for our method was about 70\% (i.e., total number of spectra sampled was $\approx 142,000$), whereas that for the \textsc{aps} was about 50\% (i.e., total number of spectra sampled was $\approx 200,000$).

\section{Discussion}
\label{discussion}

It is important to note that while our fit and the confidence set around it are nonparametric,
we do need to resort to parametric assumptions implied by the use of \textsc{camb}
for the purpose of making inference about cosmological parameters.
For example, one such parametric assumption is that of a power-law form for the primordial power spectrum characterized in terms of the parameters $n_s$ and $A_s$.
We have also taken the massive neutrino component $\Omega_\nu=0$ throughout this work.
However, neither our analysis nor our method is restricted by these parametric assumptions.

\begin{figure}
 \centering
 \includegraphics[width=0.9\textwidth]{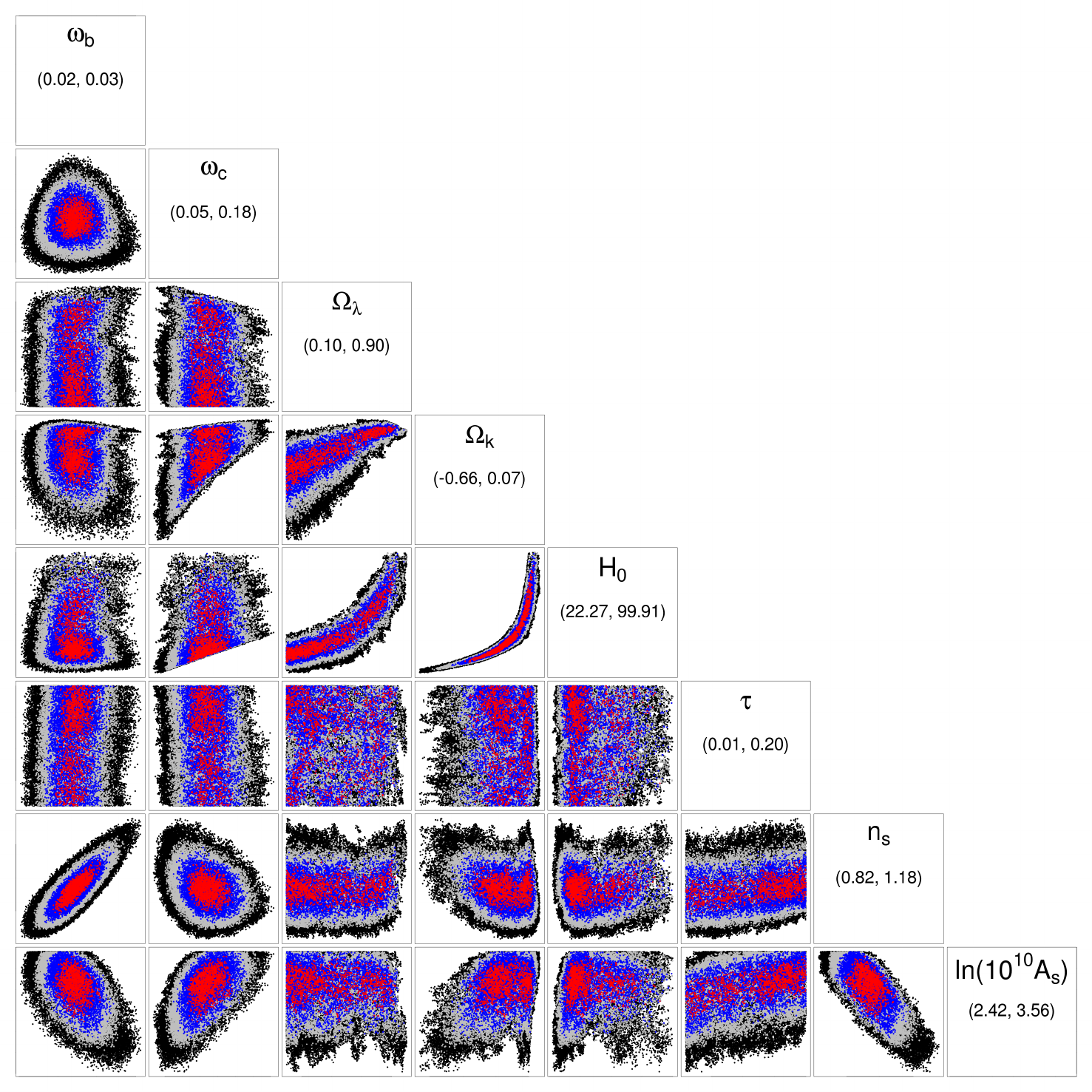}
 \caption{\label{fig:prior-nill} Nonparametric pairwise scatterplots (sample size: 85,000) of cosmological parameters at four levels of confidence; namely, 95\% ($2\sigma$, black), 68\% ($1\sigma$, gray), 38\% ($0.5\sigma$, blue), 20\% ($0.25\sigma$, red). $\Omega_k$ is a derived parameter here. Well-known \textsc{cmb}-related degeneracies and correlations in the parameters, e.g., correlations between the $(\Omega_k , H_0)$ and $(\Omega_k, \Omega_\Lambda)$ pairs, are correctly reproduced by our method. Panels on the diagonal list the range of values for the corresponding parameter.}
\end{figure}

\begin{figure}
 \centering
 \begin{minipage}{0.48\textwidth}
 \includegraphics[width=0.925\textwidth]{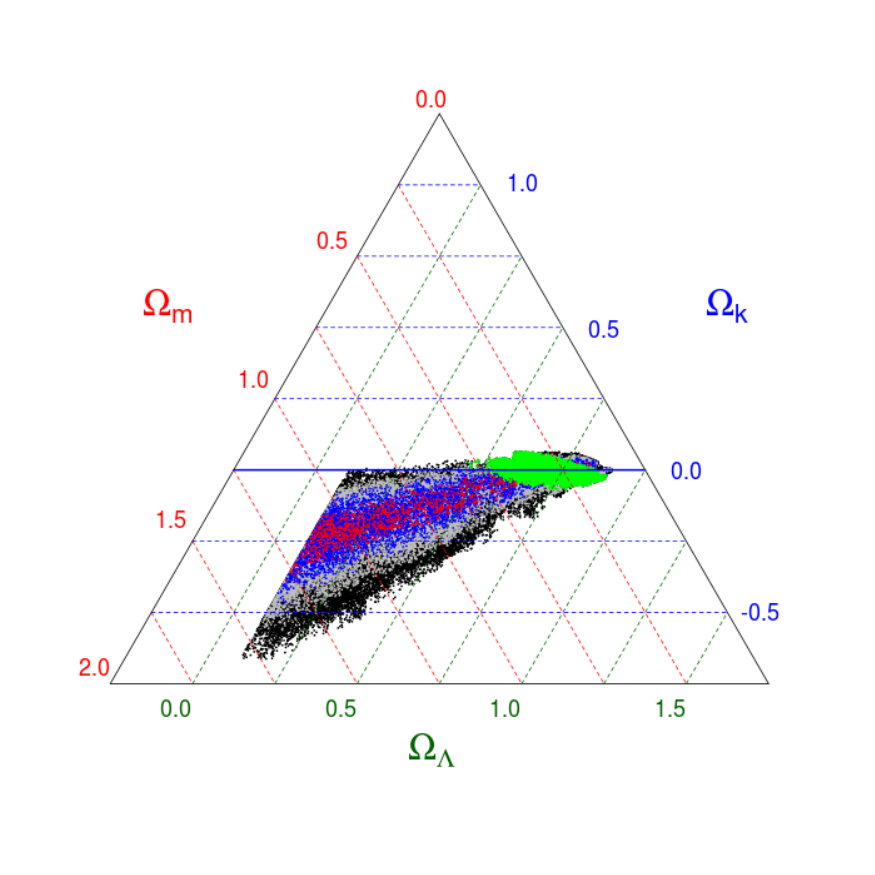}
 \end{minipage}
 \begin{minipage}{0.48\textwidth}
 \vspace{-3em}
 \includegraphics[width=0.8\textwidth]{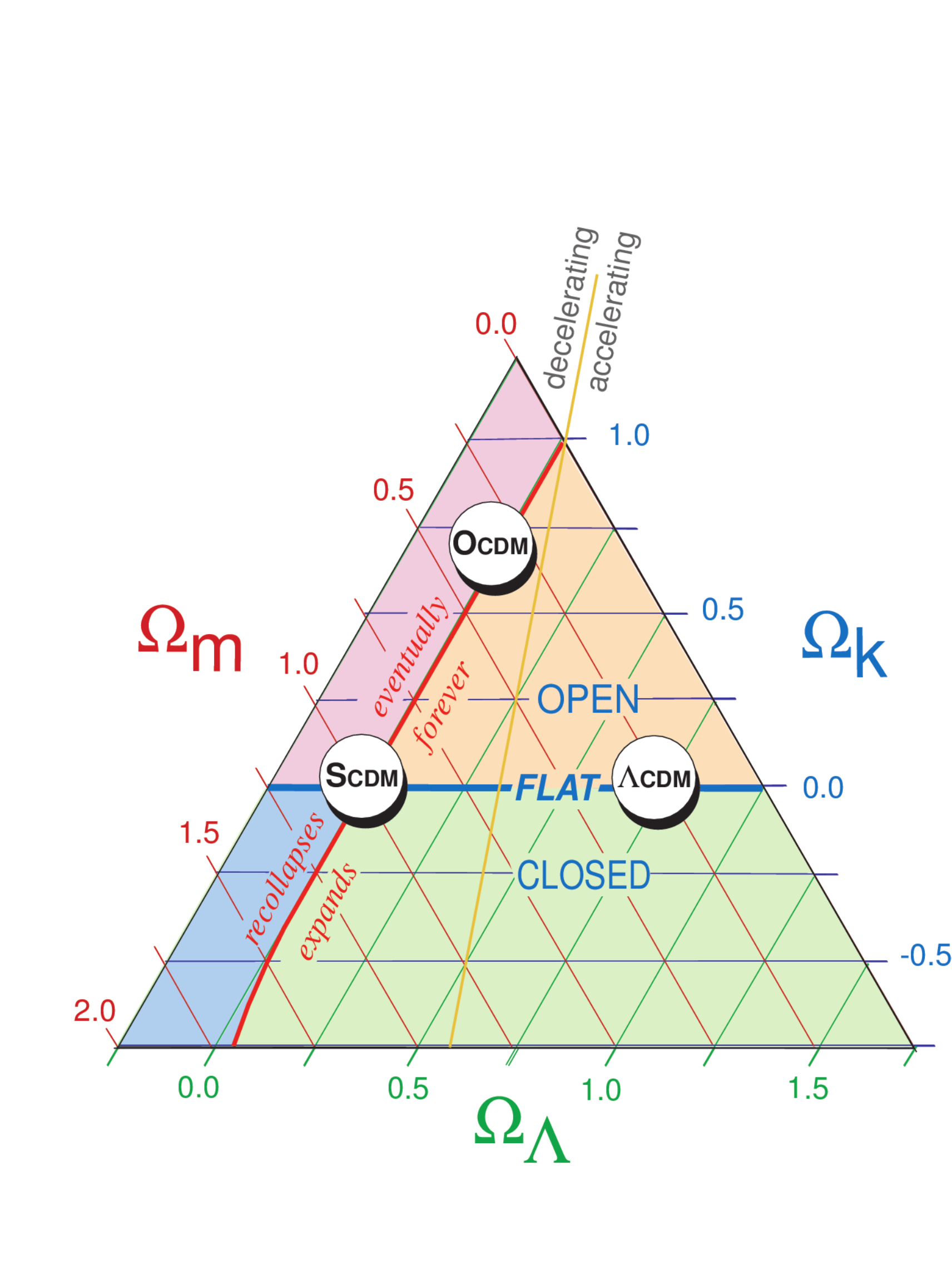}
 \end{minipage}
 \caption{\label{fig:cosmic-triangle} The nonparametric cosmic triangle plot. The right-hand panel (taken from \citet{BOP+1999}) is a legend indicating possible cosmologies in different parts of the cosmic triangle. The green scatter is for the 95\% ($2\sigma$) confidence level, but with the restriction $60 \le H_0 \le 80$. Rest of the color convention is same as that in Fig.\ \ref{fig:prior-nill}.}
\end{figure}

Our nonparametric results (Fig.\ \ref{fig:prior-nill}) suggest that uncertainties on cosmological parameters as estimated from the \textsc{wmap} 7-year data are much larger than what parametric results such as \citep{LDH+2011} imply.
Given that parametric results, which are often portrayed as ``established beyond doubt", this raises some concern.
For example, an implication of our results is that the cosmologies in those parts of the nonparametric confidence set that are not covered by parametric analyses due to inherent assumptions or limited dimensionality of the parametrized models used may not be entirely irrelevant.
Our results therefore suggest interpreting parametric confidence intervals or Bayesian credible intervals reported in literature with adequate caution.
We also note that the interpretation of frequentist confidence intervals (such as those constructed from a nonparametric confidence set) is entirely different from that of Bayesian credible intervals reported by most mainstream methods.
A thorough discussion about this can be found in (\citet{BSM+2007}; see also \citet{ABB+2002}).

Additional information (e.g., in the form of tighter constraints on $H_0$ or $\Omega_k$) combined with our nonparametric analysis leads to considerably tighter confidence intervals that are consistent with the mainstream parametric results.
We interpret this as showing that with additional information, parametric results are generally validated by our data-driven semi-parametric analysis.

Why are the nonparametric uncertainties that we report larger than those reported by model-based parametric analyses?
Parametric analyses use models that are built upon cosmological assumptions and use prior information,
whereas nonparametric inference is motivated to make the best of the data with as few assumptions as possible.
Indeed, the essential assumptions underlying our nonparametric regression methodology are about the nature of noise in the data (mean-0 normal noise with known covariances), and about the smoothness of the true but unknown angular power spectrum (that it is a square-integrable function with pre-specified effective degrees of freedom; see \citet{AAS2012}).
Less assumptions implies being more data-driven, albeit with greater uncertainty -- this is the price to pay for going nonparametric \citep{Wasserman2006}.

On the other hand, in the nonparametric regression methodology (\citet{AAS2012} and references therein) that we use, the fit is obtained by minimizing a risk function that can be expressed as the squared bias of the fit plus its variance.
Most mainstream methodologies for estimating of the unknown regression function place greater emphasis on the estimator being unbiased, and not necessarily on minimizing the complete risk.
Lower risk of this nonparametric fit implies that it is, on an average, closer to the true but unknown regression function than any unbiased estimator \citep{Beran2001}.

In summary, in this paper, we have developed a Markov chain Monte Carlo method for sampling the cosmological parameter space in such a way that the density in cosmologically meaningful regions of a nonparametric confidence set is, in principle, uniform.
This confidence set is centered at a model-independent, nonparametric fit for the \textsc{wmap} 7-year angular power spectrum data.
Our results indicate that uncertainties on cosmological parameters are much wider than the parametric confidence or credible intervals reported in literature.
Since nonparametric methods are data-driven and model-independent by nature,
this suggests using and interpreting parametric results with caution.
With additional information, however, the nonparametric uncertainties shrink considerably, thereby validating mainstream parametric results.
Specifically, our nonparametric results clearly rule out an open universe; however, validating the $\Lambda$CDM model needs additional assumptions or prior information about parameters such as $H_0$ and $\Omega_k$.
Measurements of CMB polarisation spectra are improving, and are expected to provide yet another interesting dataset for analysis using nonparametric methods.
It is also expected that additional breakdown of degeneracies in the cosmological parameters as inferred from CMB data alone will come from measurements of the weak lensing effect in the CMB at large multipoles, from experiments such as SPT \citep{RAC+2004} and ACT \citep{FAA+2010}.
We expect such data to lead to more definitive answers about cosmological parameters and models of the universe.

\acknowledgments
A.A.\ would like thank Amir Hajian for many discussions and suggestions.
M.A.\ would like thank Rajaram Nityanand for sharing his methodological insights.
T.S.\ acknowledges support from the DST Swarnajayanti Fellowship.

\bibliographystyle{apj}
\bibliography{cmb,cosmomc}

\clearpage

\begin{table}
 \centering
 {\sf \scriptsize
\begin{tabular}{c|c|c|c}
 Parameter & Without Additional Priors & $60\le H_0 \le 80 $ & $-0.005 \le \Omega_k \le +0.005$\\

 \hline
 \hline

 $\omega_b$ & & &\\
 & & & \\
 95\% & $(\sci{1.5221}{-2}, \sci{3.0604}{-2})$ & $(\sci{1.6182}{-2}, \sci{3.0210}{-2})$ & $(\sci{1.6527}{-2}, \sci{3.0249}{-2})$\\
 68\% & $(\sci{1.6875}{-2}, \sci{2.8700}{-2})$ & $(\sci{1.7294}{-2}, \sci{2.8554}{-2})$ & $(\sci{1.7609}{-2}, \sci{2.7660}{-2})$\\
 38\% & $(\sci{1.8218}{-2}, \sci{2.7128}{-2})$ & $(\sci{1.8766}{-2}, \sci{2.6617}{-2})$ & $(\sci{1.8813}{-2}, \sci{2.6770}{-2})$\\
 20\% & $(\sci{1.9423}{-2}, \sci{2.5437}{-2})$ & $(\sci{1.9730}{-2}, \sci{2.5431}{-2})$ & $(\sci{1.9730}{-2}, \sci{2.5437}{-2})$\\

 \hline

 $\omega_c$ & & &\\
 & & & \\
 95\% & $(\sci{4.7388}{-2}, \sci{1.8339}{-1})$ & $(\sci{5.0750}{-2}, \sci{1.8100}{-1})$ & $(\sci{6.4593}{-2}, \sci{1.7664}{-1})$\\
 68\% & $(\sci{6.0149}{-2}, \sci{1.6918}{-1})$ & $(\sci{6.0808}{-2}, \sci{1.6356}{-1})$ & $(\sci{6.7943}{-2}, \sci{1.6287}{-1})$\\
 38\% & $(\sci{7.0158}{-2}, \sci{1.5354}{-1})$ & $(\sci{7.0158}{-2}, \sci{1.5029}{-1})$ & $(\sci{7.6996}{-2}, \sci{1.4688}{-1})$\\
 20\% & $(\sci{8.4619}{-2}, \sci{1.3698}{-1})$ & $(\sci{9.0208}{-2}, \sci{1.3607}{-1})$ & $(\sci{9.3651}{-2}, \sci{1.3607}{-1})$\\

 \hline

 $\Omega_\Lambda$ & & &\\
 & & & \\
 95\% & $(\sci{1.0000}{-1}, \sci{8.9978}{-1})$ & $(\sci{4.4981}{-1}, \sci{8.9408}{-1})$ & $(\sci{1.0274}{-1}, \sci{8.9978}{-1})$\\
 68\% & $(\sci{1.0001}{-1}, \sci{8.9851}{-1})$ & $(\sci{5.1323}{-1}, \sci{8.7734}{-1})$ & $(\sci{3.3100}{-1}, \sci{8.9851}{-1})$\\
 38\% & $(\sci{1.0008}{-1}, \sci{8.8219}{-1})$ & $(\sci{5.4017}{-1}, \sci{8.3626}{-1})$ & $(\sci{4.3552}{-1}, \sci{8.8219}{-1})$\\
 20\% & $(\sci{1.0014}{-1}, \sci{8.2285}{-1})$ & $(\sci{5.7924}{-1}, \sci{8.0836}{-1})$ & $(\sci{5.7924}{-1}, \sci{8.0836}{-1})$\\

 \hline

 $\Omega_k$ & & & \\
 & & & \\
 95\% & $(\sci{-6.6279}{-1}, \sci{6.5572}{-2})$ & $(\sci{-6.7194}{-2}, \sci{6.5572}{-2})$ & -- \\ 
 68\% & $(\sci{-5.8463}{-1}, \sci{5.4546}{-2})$ & $(\sci{-5.5552}{-2}, \sci{5.0163}{-2})$ & -- \\ 
 38\% & $(\sci{-4.6611}{-1}, \sci{4.2133}{-2})$ & $(\sci{-4.2988}{-2}, \sci{3.5121}{-2})$ & -- \\ 
 20\% & $(\sci{-3.8682}{-1}, \sci{2.8285}{-2})$ & $(\sci{-3.5169}{-2}, \sci{2.1135}{-2})$ & -- \\ 

 \hline

 $H_0$ & & &\\
 & & & \\
 95\% & $(\sci{2.2266}{1}, \sci{9.9907}{1})$ & -- & $(\sci{4.5479}{1}, \sci{9.9502}{1})$\\
 68\% & $(\sci{2.4073}{1}, \sci{9.9712}{1})$ & -- & $(\sci{5.0354}{1}, \sci{9.8097}{1})$\\
 38\% & $(\sci{2.6807}{1}, \sci{9.9254}{1})$ & -- & $(\sci{5.4344}{1}, \sci{9.4111}{1})$\\
 20\% & $(\sci{2.9220}{1}, \sci{9.1571}{1})$ & -- & $(\sci{6.0525}{1}, \sci{8.0682}{1})$\\

 \hline

 $\tau$ & & &\\
 & & & \\
 95\% & $(\sci{1.0004}{-2}, \sci{2.0000}{-1})$ & $(\sci{1.0015}{-2}, \sci{1.9999}{-1})$ & $(\sci{1.0055}{-2}, \sci{1.9985}{-1})$\\
 68\% & $(\sci{1.0004}{-2}, \sci{1.9999}{-1})$ & $(\sci{1.0037}{-2}, \sci{1.9998}{-1})$ & $(\sci{1.0678}{-2}, \sci{1.9985}{-1})$\\
 38\% & $(\sci{1.0037}{-2}, \sci{1.9999}{-1})$ & $(\sci{1.0037}{-2}, \sci{1.9985}{-1})$ & $(\sci{1.1057}{-2}, \sci{1.9985}{-1})$\\
 20\% & $(\sci{1.0037}{-2}, \sci{1.9980}{-1})$ & $(\sci{1.0037}{-2}, \sci{1.9932}{-1})$ & $(\sci{3.6789}{-2}, \sci{1.9932}{-1})$\\

 \hline

 $n_s$ & & &\\
 & & & \\
 95\% & $(\sci{8.1640}{-1}, \sci{1.1769}{0})$ & $(\sci{8.2166}{-1}, \sci{1.1538}{0})$ & $(\sci{8.2166}{-1}, \sci{1.1729}{0})$\\
 68\% & $(\sci{8.4429}{-1}, \sci{1.1293}{0})$ & $(\sci{8.4869}{-1}, \sci{1.1043}{0})$ & $(\sci{8.4869}{-1}, \sci{1.1181}{0})$\\
 38\% & $(\sci{8.6566}{-1}, \sci{1.0906}{0})$ & $(\sci{8.7316}{-1}, \sci{1.0831}{0})$ & $(\sci{8.7681}{-1}, \sci{1.0906}{0})$\\
 20\% & $(\sci{8.9142}{-1}, \sci{1.0508}{0})$ & $(\sci{9.0707}{-1}, \sci{1.0398}{0})$ & $(\sci{9.0759}{-1}, \sci{1.0436}{0})$\\

 \hline

 $\ln(10^{10} A_s)$ & & &\\
 & & & \\
 95\% & $(\sci{2.4179}{0}, \sci{3.5553}{0})$ & $(\sci{2.4429}{0}, \sci{3.5553}{0})$ & $(\sci{2.4179}{0}, \sci{3.5553}{0})$\\
 68\% & $(\sci{2.6225}{0}, \sci{3.5553}{0})$ & $(\sci{2.6225}{0}, \sci{3.5541}{0})$ & $(\sci{2.7164}{0}, \sci{3.5522}{0})$\\
 38\% & $(\sci{2.7442}{0}, \sci{3.5550}{0})$ & $(\sci{2.8879}{0}, \sci{3.5508}{0})$ & $(\sci{2.8505}{0}, \sci{3.5508}{0})$\\
 20\% & $(\sci{2.8938}{0}, \sci{3.5103}{0})$ & $(\sci{2.9915}{0}, \sci{3.4520}{0})$ & $(\sci{3.0317}{0}, \sci{3.3991}{0})$\\
\end{tabular}
 }
 \caption{\label{tab:ci} Nonparametric confidence intervals on cosmological parameters at 95\% ($2\sigma$), 68\% ($1\sigma$), 38\% ($0.5\sigma$), 20\% ($0.25\sigma$) confidence levels. Column 1 lists nonparametric confidence intervals without any additional information. Column 2 and 3 list nonparametric confidence intervals after constraining $H_0$ and $\Omega_k$ respectively.}
\end{table}

\clearpage

\begin{table}
 \centering
 {\sf \small
\begin{tabular}{l|c|c}
 Parameter          & Extreme Values in the Parametric Scatter & Nonparametric 95\% Confidence Interval \\
                    & $(\Omega_k=0)$                            & Subject to $-0.005 \le \Omega_k \le +0.005$ \\
 \hline \hline
 $\omega_b$         & $(0.020364, 0.024994)$ & $(0.016527, 0.030249)$\\
 $\omega_c$         & $(0.091191, 0.13275)$ & $(0.064593, 0.17664)$\\
 $\Omega_\Lambda$   & $(0.59986, 0.82709)$ & $(0.10274, 0.89978)$\\
 $H_0$              & $(61.911, 82.318)$   & $(45.479, 99.502)$\\
 $\tau$             & $(0.036069, 0.16104)$ & $(0.010055, 0.19985)$\\
 $n_s$              & $(0.90402, 1.0181)$  & $(0.82166, 1.1729)$\\
 $\ln(10^{10} A_s)$ & $(3.01232, 3.3726)$  & $(2.4179, 3.5553)$\\
\end{tabular}
 }
 \caption{\label{tab:pe} Extreme parameter values in the parametric scatter (red) in Fig.\ \ref{fig:p-vs-np}. This parametric scatter assumes a flat universe ($\Omega_k=0$). For comparison, we have also included our nonparametric 95\% confidence intervals subject to $-0.005 \le \Omega_k \le 0.005$.}
\end{table}

\clearpage

\clearpage

\clearpage

\clearpage

\clearpage

\clearpage

\clearpage

\clearpage

\clearpage

\end{document}